\documentclass[reprint,amsmath,amssymb,nofootinbib,aps]{revtex4-2}
\usepackage[english]{babel}
\usepackage[utf8]{inputenc}
\usepackage{adjustbox}
\usepackage{afterpage}
\usepackage{siunitx}
\usepackage{float}
\usepackage{xcolor}

\usepackage[twocolumn,textwidth=16.6cm,columnsep=.61cm]{geometry}

\begin{document}
\title{Transport of soft matter in complex and confined environments}
\author{Joshua D. McGraw$^{1,2}$}
\email{joshua.mcgraw@espci.psl.eu}

\affiliation{$^{1}$Gulliver UMR 7083 CNRS, ESPCI--PSL, 10 rue Vauquelin, 75005 Paris, France}
\affiliation{$^{2}$IPGG, 6 rue Jean-Calvin, 75005 Paris, France}

\date{\today} 


\begin{abstract}
Brownian motion provides a bedrock for the understanding of soft condensed matter and, therefore, of the physical description of the microscopic biological world. Inspired by this domain, and combining softness with hydrodynamic energy inputs, new physical modes of nanoscale organization and transport may now be accessible. 
\end{abstract}


\maketitle


Diffusion and interfacial interactions are at the root of a huge number of fundamental structural and transport processes in complex environments. A salient example of a such an environment is that of cellular and sub-cellular biology: these systems are mixtures containing biomacromolecules such as proteins and DNA, with soft interfaces including cell walls, membranes and polymer brush-like coatings never too far away. From a physical standpoint, these interfacial systems are primarily built from a selection of soft matter building blocks near soft boundaries. Another physical feature of these environments is that they are out of equilibrium, often due to an imposed flow. This hydrodynamics implies viscous dissipation and so a constant energy input. Such flows can generate non-conservative (i.e., velocity-dependent) lift forces between soft particles and their surroundings. These interactions can be comparable to thermally-mediated, potential-derived ones and thus may provide original levers for organization and transport out of equilibrium.  

In many physical contexts, a basic question is that of where a given thing will be at a given instant. In soft matter science, the answer to this kind of question is typically statistical in nature, owing to the fact that the associated transport is dominated by thermal fluctuations. Indeed, \textbf{understanding the transport of soft matter is often dependent on the interpretation of Brownian motion and its collective consequences over diverse length and time scales}. This motion and related diffusion problems have been treated for many decades~\cite{Haw2002, Duplantier2005} by the now standard theories of Boltzmann, Einstein and Stokes, which have been extremely successful in describing many microscopic systems. Respectively, these three provide: an eponymous probability distribution describing the spatial organization of soft matter; a diffusion coefficient, depending on thermal fluctuations and friction through the fluctuation-dissipation relation; and, a continuum statement for the overdamped conservation of momentum. Coupling these, transport equations involving soft matter players is the bedrock for physical descriptions of statistical evolutions, typically through advection-diffusion-like equations for time-dependent concentration fields or probability distributions. 

In the remainder of this short primer, we first highlight the general importance of Brownian motion in soft matter. Then, after mentioning the classic potential-based strategy for self-organization, we discuss some relatively new and promising forms of particle manipulation based on hydrodynamic interactions with a confining interface. In closing, we make some remarks on how this organization could influence transport in dense interfacial and complex systems. While mainly speculative, the concluding discussion echoes: a recent call to put transport at the focus of related micro- and nanoscale wetting phenomenology, for simple fluids~\cite{Bocquet2011}; and a similar one to investigate general complex materials by France’s recent national prospective on physics~\cite{Restagno2024}. 

\subsection*{Brownian motion and \\ Boltzmann statistics at equilibrium}

\noindent Considering the object that launched the soft matter field~\cite{Haw2002, Duplantier2005}, a colloid is essentially defined by its susceptibility to the thermal fluctuations (small parallel lines throughout Fig. 1) of surrounding molecules. Such an object has length scales ranging from a few nanometers and up to a couple of micrometers. Colloidal objects —such as the pollen, and, more pointedly, the inorganic grains that were both observed by Brown— constantly move erratically. This erratic motion in the form of a random walk underpins without too much exaggeration all of soft matter science. 

\begin{figure}[t!]
\centering
\includegraphics[width=\columnwidth]{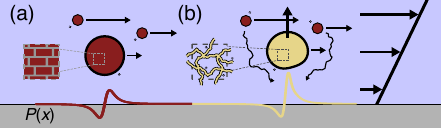} 
\caption{Transport of (a) rigid and (b) soft colloidal-scale objects near interfaces by a shear flowing viscous liquid (profile at left). The anti-symmetric and symmetry-broken pressure fields, $P(x)$, are shown schematically, the latter generating lift (vertical arrow) of the soft particle. }
\label{fig:Tranport}
\end{figure}

To provide concrete illustrations of Brownian motion’s importance to soft matter, we consider another important building block of the field. The simplest polymer molecule has a chain-like structure with many repeating, identical chemical units, as in Fig. 1(b) inset.\footnote{In this representation, the individual linear molecules are part of a network making a soft microgel particle. } Thousands of different chemical species are possible, while the aforementioned protein and DNA molecules are heterogeneous versions of the same linear structure. When homogeneous polymer molecules are the only species in a material, an individual molecule takes, statistically, the shape of an uncorrelated random walk, akin to the trajectory of a colloid in a viscous bath. Enumeration of the configurations a molecule can take at a given size endow the molecules with the property of a Hookean spring. The spring constant has an entropic origin, the molecule is thus considered soft.

Starting from the shape of a polymer chain and its entropic springiness, it is possible to build microscopic dynamical models of polymer molecules by considering the balance between viscous and thermal stresses, through the Einstein relation, and considering the random-walking structure of the molecule. The Rouse and entanglement models, familiar to polymer physicists, capture the main features of rheological spectra, containing storage and loss components, of polymeric materials. The statistical answer to the quite basic question “where are the polymer molecules?” thus informs industrial engineers, for example, in the field of polymer processing. The energy input required of the macroscopic processing machines is intimately linked with the microscopic random walks of the molecules and their temporally diffusive explorations. If such a statement is relevant for bulk systems containing only polymer molecules, so it must also be for heterogeneous interfacial systems as well. 

Soft matter indeed offers a playground of interfacial interactions that tailor spatial structure and dynamics of micro- and meso-scale systems. For example, in the aqueous environments that are common to biophysical environments, charged surfaces (one of a colloidal particle, say, or a charged polymer) can repel one another through an exponentially-decaying potential interaction characterized by the Debye length of nanometric scale. Using standard theories of statistical mechanics, the distribution of particles above a surface can thus be predicted; the Boltzmann distribution is emblematic. A similar distribution could be observed considering a gravitational or van der Waals interaction. All of these examples, since they are governed by potentials, depend only on the instantaneous location of the particle or configuration of an ensemble; i.e., in Fig. 1, the distribution would be insensitive to the particle velocities (horizontal arrows). However, \textbf{standard self-organization and transport scenarios may be greatly perturbed by surrounding flows, especially when the objects being transported, or the boundaries of the flow, are soft.}  

\subsection*{Elastohydrodynamic interactions under flow}

Indeed, over the last couple of decades, an elastohydrodynamic lift paradigm has been intensely investigated~\cite{Bureau2023, Rallabandi2024a}, wherein a particle transported near a boundary in a viscous fluid experiences a separation-dependent repulsion when either the particle or surface is deformable. Briefly, this lift arises because the viscous stresses of the surrounding fluid modify the shape of the elastic interface, as in Fig. 1(b). This modification renders the hydrodynamic pressure field in the fluid around the particle asymmetric, as shown schematically by the two profiles, P(x), in Fig. 1. The integral of the pressure over the surface, i.e. the lift force, is thus non-zero for the soft particle. Detailed descriptions also depend on the local flow field, viscosity, and particle size. 

Besides an elastic wall or particle, it turns out that any kind of asymmetry that can be induced by a flow may generate lift forces of the kind described above, and any kind of softness, too. To name a few more, electrokinetic effects due to flow-induced deformation of the equilibrium electrolyte cloud (itself diffusively mediated) in the above-mentioned electrolytic context, or crossing a heterogeneity in wall slip —due to a heterogeneous surface, say— can also generate lift forces on particles. Furthermore, the surface does not need to be purely elastic; surface tension between a liquid-liquid interface also maintains the feature of deformability. 

Coming back to the key, soft-matter player evoked above, dilute polymer molecules in solution also exhibit large modifications to the average concentration profile~\cite{Graham2011} in near-surface geometries under flow. The modifications are governed by hydrodynamic interactions with the boundary coupled to the random-walking shape and softness of the polymer chain. For such a nanoscale object, the hydrodynamically mediated wall-normal fluxes compete with a diffusive one. In this way, a situation in which non-equilibrium, hydrodynamic interactions modify the probability distribution is encountered. Brownian elastohydrodynamics~\cite{Fares2024} may thus be seen as an effective physico-chemical lever —independent from e.g. the salt concentration for controlling electrostatic interactions— useful for the control of spatial organization in dense soft matter systems. Next, we speculate as to how this might work in complex systems, and finally make a comment about advective transport in dense systems, returning to the central question of where to find objects. 

\subsection*{Segregation and transport in heterogeneous soft matter}

The complex scenario described in the first paragraph involved mixtures of macromolecular and other colloidal-scale objects near surfaces and subject to flow. We established that such confined flows generate wall repulsions of hydrodynamic origin. Considering the multi-component and crowded nature of a complex mixture, the common flow rate has a strong potential to alter the balance of diffusive and potential-derived fluxes that would otherwise be captured by a Boltzmann distribution. In this heterogeneous, flowing context, each species (i.e., macromolecule or colloidal object) is subject to a different repulsive force at a given distance from the wall. 

In such a flowing heterogeneous medium, an interesting potential feature is the creation of an “effective” attractive interaction, even while induced-hydrodynamic interactions with a surface are typically repulsive. This could happen in a suspension containing some rigid and some soft objects, for example. While soft objects are repulsed during a flow, an initially homogeneous distribution of the different species could preferentially allow the rigid objects to explore the newly exposed areas near the surface as the soft ones are repelled into the bulk, as in Fig. 1(b). This “effective” attraction hypothesis is reminiscent, albeit with a sign inversion, of an interesting collective result concerning van der Waals interactions. All molecule pairs attract one another through orientation-averaged, dipole-dipole interactions, yet interfaces may repel one another depending on the strength of relative attractions across different materials. 

Concerning particle transport, the driving mechanism for the segregation hypothesis above is the shear flow typically observed at interfaces and indicated in Fig. 1. Thus, in addition to wall-normal segregation, different species are advected at different average rates. Meanwhile, the colloidal character of the objects leads them to explore the space around their average positions, leading to the stochastic sampling of different velocity streamlines. This is exactly the transport phenomenon that Taylor described decades ago~\cite{Taylor1953} [9] in work considering a single diffusing species in a pipe flow. This advection-diffusion coupling gives rise to enhanced particle spreading, called Taylor dispersion, much faster than molecular diffusion only. This dispersion was recently investigated for some minimally complex systems at the nanoscale~\cite{Vilquin2023}, including non-trivial interactions between particles and their confining boundaries, putting into evidence the importance of non-equilibrium distributions on near-wall transport. 

\textbf{Combining transport and Brownian elastohydrodynamic segregation principles at the nanoscale, there are many opportunities for controlling complex, heterogeneous transport in confinement.} Segregation and separation schemes based on non-equilibrium interactions in micro- or nano-fluidic devices could be devised, these reminiscent of inertial focusing at larger scales. One may also tune the distribution of softness in a mixture to prepare desired vertical concentration profiles. With the distance from the wall determining velocity, and coupled to a tailoring of the vertical diffusivity with size selection, such a scheme could allow for controlled, temporal concentration profiles at prescribed distances down a channel. 

Even while highly speculative, the structure-transport interplay noted above provides exciting avenues for micro and nanoscale physics and engineering. Inspired by biological players which, at their root, are soft matter species, the schemes are thus generalizable to a broad class of objects and applications. It appears that, just as Brownian motion had as its root a biological inspiration but turned out to be a fundamentally physical phenomenon, a strain of current physics can again be greatly inspired by micro- and nano-scale biology to advance the understanding of general, soft and complex systems. 

\textbf{Acknowledgements:} Collaborations with T. Salez, F. Restagno, M. Le Merrer, C. Barentin, E. Raphaël, D. Dean, M. Labousse, A. Vilquin, V. Bertin, G. Guyard, M. Gunny and the GDR ISM team contributed greatly to the development of the ideas presented here. The author warmly acknowledges our many discussions. The work also benefitted from the financial support of the ANR NoDiCE grant (ANR-24-ERCC-0003).

\textbf{Author bio:} The author is soft matter physicist working on interfacial dynamical problems including: diffusive transport at interfaces, soft hydraulics, lubrication of soft contacts with complex fluids, poroelasticity, and confined elastocapillary self-assembly.  
  

%

\end{document}